\documentclass[aps, pre, longbibliography, superscriptaddress,
notitlepage]{revtex4-2}
\usepackage{bm, amsmath, amssymb, graphicx, enumerate, mathtools, multirow, booktabs}
\usepackage{float}
\usepackage[colorlinks=true,linkcolor=MidnightBlue,urlcolor=black,
            citecolor=MidnightBlue,anchorcolor=MidnightBlue]{hyperref}
\usepackage[dvipsnames]{xcolor}

\newcommand{\avg}[1]{\langle #1 \rangle}
\newcommand{\Polv}{|\mathbf{P}|}
\newcommand{\Npc}{N_{\mathrm pc}}
\newcommand{\Nc}{N_c}
\newcommand{\phiA}{\phi}
\newcommand{\chiT}{\chi_T}
\newcommand{\chiR}{\chi_R}
\newcommand{\vs}{v_s}
\newcommand{\Jzero}{J_0}
\newcommand{\ksp}{k_{\mathrm sp}}
\newcommand{\kbend}{k_{\mathrm bend}}
\newcommand{\Lx}{L}
\newcommand{\bead}{b}
\newcommand{\Ek}{E(k)}
\newcommand{\Ok}{\Omega(k)}
\newcommand{\Cv}{C_v(r)}

\newcommand{\DeltaN}{\Delta N}
\newcommand{\Navg}{\langle N \rangle}

\newcommand{\xivec}{\bm{\xi}}
\newcommand{\nhat}{\hat{\bm{n}}}
\newcommand{\Dr}{D_r}

\begin{document}
\title{Collective dynamics of chemo-mechanical colloidal chains with active tips}

\author{Arvin Gopal Subramaniam}
\thanks{ph22d800@smail.iitm.ac.in}
\affiliation{Department of Physics, Indian Institute of Technology Madras,
             Chennai, India}

\author{Rajesh Singh}
\thanks{rsingh@physics.iitm.ac.in}
\affiliation{Department of Physics, Indian Institute of Technology Madras,
             Chennai, India}

\begin{abstract}
We report a study of the emergent dynamics arising in
two-dimensional suspensions of semi-flexible 
colloidal chains whose tip is chemically active, generating a  
phoretic field.
By varying the chain length (number of monomers per chain~$\Npc$),
the area fraction~$\phiA$, and the sign of the phoretic coupling~$\Jzero$,
we map out a rich non-equilibrium phase diagram in the presence of
phoretic interactions.
For repulsive phoretic interactions 
($\Jzero > 0$) between the chains,
we find that the shortest chains ($\Npc = 2$) develop a
global polar flock with suppressed density fluctuations.
At intermediate chain lengths ($\Npc \sim 4$--$8$), the repulsive chemical
field drives chaotic mesoscale flows --- a dry route to active turbulence
without the need to hydrodynamic interactions
or steric alignment interactions. 
A swarm phase - characterized by the coexistence of local polar order and (local) vortex pairs - is also reported and discussed.
For attractive phoretic interactions ($\Jzero < 0$), chains self-organize into
hedgehog-like micellar aggregates with heads forming the core and flexible
tails radiating outward, in structural analogy with amphiphile micellization
but driven entirely by non-equilibrium self-propulsion.
A coarse-grained theory of a tip-emitting active rods predicts the onset of the flocking of dimers, though overestimates the presence of polar order for longer chains.
Our results establish phoretic tip activity as a minimal, experimentally
realizable mechanism for a spectrum of collective states hitherto attributed
to hydrodynamic interactions or steric alignment.
\end{abstract}

\maketitle

\section{Introduction}\label{sec:intro}

Collections of self-propelled particles display a remarkable wealth of
collective behaviour with no equilibrium counterpart, including polar
flocking~\cite{Vicsek1995,toner1995long}, motility-induced phase separation
(MIPS)~\cite{cates2015motility}, active turbulence~\cite{Alert2022}, and
non-equilibrium self-assembly~\cite{Mallory2018review}.
The mechanisms driving these phases --- steric alignment, hydrodynamic
torques, short-range repulsion --- have been systematically dissected
in canonical systems of spherical active Brownian particles and rigid self-propelled
rods~\cite{Wensink2012PNAS,wensink2012emergent,Bar2020}.

A key distinction among these active constituents is whether the
symmetry-breaking that drives self-propulsion is \emph{built into} the
particle at synthesis, or arises \emph{spontaneously} from an
instability in an otherwise symmetric system.
In the first class, activity is localised by construction -- Janus
colloids with a catalytic cap on one hemisphere~\cite{Howse2007},
hematite-doped particles with an embedded photocatalytic
domain~\cite{palacci2013living}, and shape-asymmetric active particles
such as chiral granular ellipsoids~\cite{Arora2021} -- all fix the location
and direction of the active flux permanently in the particle geometry
or chemistry, independent of the surrounding dynamics.
In the second class, the particle is fabricated with no built-in
asymmetry, and directionality instead emerges dynamically: isotropic
oil or water droplets that spontaneously self-propel via a
Marangoni-stress instability once a control parameter (P\'eclet number)
exceeds a threshold~\cite{Izri2014, kumar2024emergent}, or symmetric dielectric spheres
that spontaneously begin rolling under a uniform electric field via the
Quincke instability~\cite{Bricard2013}.
The system that we study in this paper is within the first class: the active tip is a fixed, being a
permanent feature of each chain.

A separate and largely orthogonal body of work has examined collective
behaviour in active \emph{extended} objects assembled by physically
linking individually active units --- colloids, droplets, or
Janus particles --- into a chain, rather than treating the active
body itself as a single rigid, elongated shape.
Examples include catalytically-active colloids linked into flexible
chains to enhance their diffusivity~\cite{Biswas2017}, Janus
particles assembled into chains that exhibit internally-driven,
flagellum-like beating~\cite{Nishiguchi2018}, externally-reconfigurable
Janus colloids whose interactions can be tuned to chain or
cluster~\cite{Yan2016}, and
freely-jointed active polymers built from self-propelled droplet
monomers~\cite{kumar2024emergent}.
In essentially all of this work, however, activity is distributed
across every unit in the chain rather than localised to a single,
fixed end.

This work combines both strands: an active \emph{extended} object
(a flexible bead-spring chain) whose activity is spatially localised
to a single, permanently fixed tip, i.e.\ an extended body with
\emph{built-in} rather than emergent asymmetry.
This combination was recently synthesised and studied experimentally
in a system built from light-driven colloidal rods composed of a
catalytic TiO$_2$ head and a passive SiO$_2$
tail~\cite{Shelke2026Science} (Fig.~\ref{fig:phasediag}a).
Photocatalytic reactions at the head generate local chemical gradients
that propel the rod along its long axis; particle image velocimetry
confirms a pusher-type hydrodynamic stresslet whose chemical origin is
the monopolar concentration field at the tip.
The authors reported a various range of collective
states including swarming, active turbulence, flocking, and jamming, and concluded hydrodynamic
interactions are necessary to recover the experimentally observed
phases~\cite{Shelke2026Science}.

In this 
paper, we study semi-flexible chains with active tips. These semi-flexible chains only have phoretic interactions due to the active-tip. We find a rich phenomenology as the length of the chain and packing fractions is varied. 
Our results are broadly classified for active-tips with attractive and repulsive phoretic interactions. 
For a chain of aspect ratio 2, we find polar phases with suppressed density fluctuations.  Thus, indicating the emergence of hyperuniformity \cite{torquato2016hyperuniformity, subramaniam2026shape, cates2025active}. On increasing the aspect ratio, we find a dry route to active turbulence. Notably, this turbulent state is obtained in absence of any hydrodynamic interactions. Further, we also report swarming states - characterized by local polar order global spatial disorder. 
In the case of attractive phoretic interactions mediated by the active-tip of the chains, we find distinct steady-states of packed flexible chains including hedgehog-like micellar aggregates with heads forming the core and flexible
tails radiating outward. This state has a
structural analogy with amphiphile micellisation
but driven entirely by non-equilibrium phoretic interactions.
We also find liposomal and glassy states for attractive interactions, which we characterize in detail. 
A hydrodynamic description has also been constructed by systematically coarse-graining a microscopic model of tip-active rigid chains. 
The hydrodynamic theory captures the flocking transition for dimers, but overestimates the existence of the polar instability for longer chains.

The rest of the 
paper is organized as follows.
Section~\ref{sec:model} describes the model.
Sections~\ref{sec:repulsive} and~\ref{sec:attractive} present the results for the cases  of
repulsive and attractive phoretic interactions respectively.
In addition, in section \ref{sec:theory} of the paper, we present a continuum theory to rationalize our numerical results. 
Section~\ref{sec:discussion} summarizes the result and places them in broader context.

\section{Model and methods}\label{sec:model}

\subsection{Chain architecture and equations of motion}

\begin{figure}[t!]
  \centering
  \includegraphics[width=0.8\textwidth]{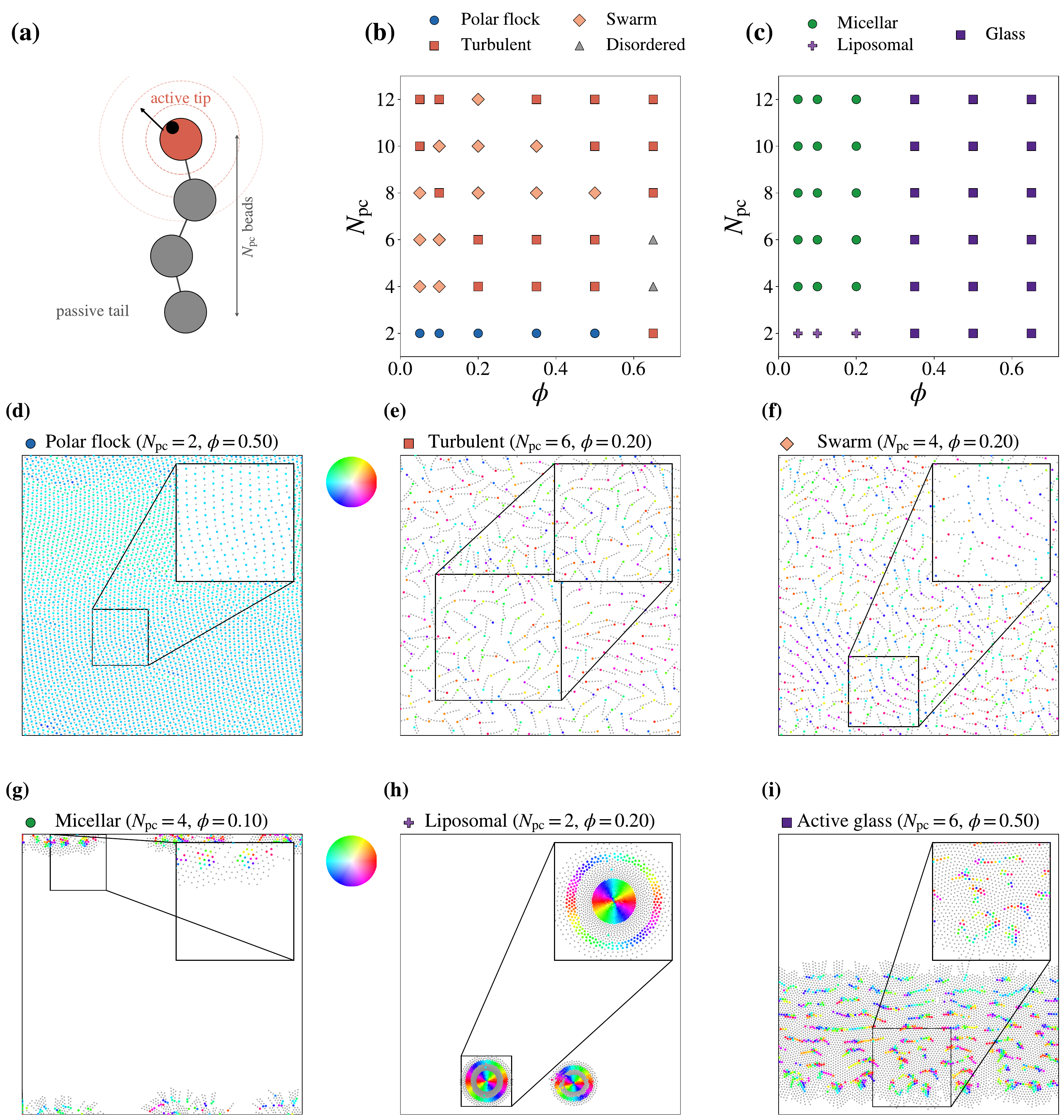}
  \caption{%
  \textbf{Model schematic, phase diagrams, and representative phases.}
  (a) A bead-spring chain of $\Npc$ monomers (grey) with a chemically
  active head (red). The head emits a monopolar concentration field
  $c \sim J_0/r$ (dashed circles) and self-propels at speed $\vs$ (arrow).
  For $J_0 > 0$ the phoretic interaction is repulsive (heads push apart
  and reorient away from each other); for $J_0 < 0$ it is attractive.
  (b) Phase diagram in ($\phi$, $\Npc$) space for repulsive phoresis
  ($J_0 > 0$): polar flock (circles), active turbulence (squares),
  swarm (diamonds), and disordered gas (triangles). Active turbulence
  occupies a narrow window that closes with increasing $\Npc$; all
  $\Npc \geq 10$ points, and the remaining $\Npc = 8$ densities, are the
  swarm phase.
  (c) Attractive phoretic interactions ($J_0 < 0$): micellar phase (circles,
  $\Npc \geq 4$), liposomal/stacked phase (plus symbols, $\Npc = 2$
  only), and active glass (squares) --- a dense, dynamically-arrested,
  locally-ordered packing.
  (d)--(f) Representative repulsive examples --- polar flock
  ($\Npc = 2$, $\phi = 0.50$), active turbulence ($\Npc = 8$,
  $\phi = 0.20$), and swarm ($\Npc = 4$, $\phi = 0.20$) --- each with a
  zoom inset on the region of interest.
  (g)--(i) Representative attractive examples --- micellar
  ($\Npc = 4$, $\phi = 0.10$), liposomal ($\Npc = 2$, $\phi = 0.20$),
  and active glass ($\Npc = 6$, $\phi = 0.50$) --- each with a zoom
  inset centered on a representative aggregate.
  We note that our coloring convention colors the head of the chain according to the color wheels, with the remaining monomers colored in grey, indicating no inherent activity.
  All symbols are steady-state observations from the simulations
  described in Sec.~\ref{sec:model}.
}
  \label{fig:phasediag}
\end{figure}

We simulate $\Nc$ bead-spring chains, each consisting of $\Npc$ monomers
of radius~$\bead$, in a two-dimensional square periodic box of side~$\Lx$.
The total number of monomers is $N = \Nc \Npc$.
Each active filament 
has a head monomer (index $i \equiv 0\ {\mathrm mod}\ \Npc$)
that carries chemical activity, and $\Npc - 1$ passive body monomers
(Fig.~\ref{fig:phasediag}a).
The overdamped equations of motion are
\begin{subequations}\label{eq:eom}
\begin{align}
    \dot{\mathbf{r}}_i &= \vs\,\hat{\bm e}_i\,\delta_{i\,{\mathrm mod}\,\Npc,\,0}
        + \mu\mathbf{F}_i^{\mathrm mech}
        + \chiT\,\mathbf{J}_i
        + \sqrt{2D_t}\,\xi_i(t), \label{eq:rdot}\\
    \dot{\theta}_i &= \chiR\,(\hat{\bm e}_i \times \mathbf{J}_i)_z
        + \sqrt{2\Dr}\,\eta_i(t), \label{eq:thdot}
\end{align}
\end{subequations}
where $\vs$ is the self-propulsion speed (head only),
$\mu = 1/(6\pi\bead)$ is the translational mobility,
$\mathbf{F}_i^{\mathrm mech}$ is the total mechanical force (springs,
bending rigidity, excluded volume),
$\chiT$ and $\chiR$ are translational and rotational phoretic couplings,
$\xi_i(t)$ and $\eta_i(t)$ are unit-variance white noise, and
$\mathbf{J}_i$ is the phoretic flux at monomer~$i$,
\begin{equation}\label{eq:Jev}
    \mathbf{J}_i = \Jzero \sum_{\substack{j \neq i \\ j\,{\mathrm mod}\,\Npc=0}}
    \frac{\mathbf{r}_i - \mathbf{r}_j}{|\mathbf{r}_i - \mathbf{r}_j|^3}.
\end{equation}
The sign of $\Jzero$ controls the character of the chemical interaction:
$\Jzero > 0$ (repulsion) or $\Jzero < 0$ (attraction).
The mechanical forces are: harmonic springs between consecutive monomers
(rest length $\ell_0 = 2\bead$, stiffness $\ksp$), a bending rigidity
$\kbend$ penalizing changes in bond angle, and a Hookean excluded-volume
repulsion active for $r_{ij} < \ell_0$.
We note that, for the whole of the paper, $D_t = \Dr = 0$ (deterministic simulation) unless stated otherwise.

\subsection{Order parameters}
We characterise collective behaviour using the global polar order
$\Polv =\frac1{ N_c}\left| \sum_j \hat{\bm e}_j \right|.$ 
In addition, to delineate phases, we use
mean-squared displacement of
head monomers; the angle-averaged energy spectrum $\Ek$ and enstrophy
spectrum $\Ok$ from a Poissonian-gridded velocity field; the velocity
correlation $\Cv$; and number
fluctuations $\DeltaN$ vs $\Navg$ over box subdivisions
$n \in \{2, 4, 8, 16, 32\}$. Expressions of these quantities and simulation details is given in appendix \ref{app:methods}, along with details of methodolgy used to obtain the results of this paper.

\section{Repulsive phoretic interactions}\label{sec:repulsive}
In this section, we consider the emergent dynamics of colloidal chains with active-tips that mediate repulsive phoretic interactions ( $J_0 > 0$). 
Results are summarized in Fig.~\ref{fig:phasediag}(b), (d)--(f).
Notably, four distinct steady-state regimes are identified in this regime as a function of aspect ratio (or the number of particles $\Npc$ in the filament) of the filament and the packing fraction ($\phi$). 
\begin{itemize}
  \item \emph{Polar flock} ($\Npc = 2$, $\phiA \leq 0.50$): global polar
  order $\Polv \to 1$ at long times, preceded by a transient turbulent
  phase; re-entrant disordering occurs at $\phiA = 0.65$ where crowding
  prevents alignment.
  \item \emph{Active turbulence} ($\Npc = 6$ at $\phiA = 0.15$--$0.65$,
  except $\phiA = 0.50$; $\Npc = 8$ at $\phiA = 0.10, 0.65$; $\Npc = 10$
  at $\phiA = 0.05, \geq 0.50$; $\Npc = 12$ at $\phiA \leq 
  0.20, \phiA \geq 0.50$): steady-state chaotic flow with
  $\Polv \rightarrow 0$, presence of vortices spanning the flow field, and a
  negative minimum in $\Cv$.
  \item \emph{Swarm} ($\Npc = 6$ at $\phiA = 0.10, 0.50$; $\Npc = 8$ at
  $\phiA = 0.15$--$0.50$; $\Npc = 10$ at $\phiA = 0.10$--$0.35$;
  $\Npc = 12$ at $\phiA = 0.20$): local vortex pairs coexist with sustained local polar order
  $\Polv \approx 0.2$--$0.4$ and no negative $\Cv$ minimum,
  distinguishing this phase from pure turbulence.
  \item \emph{Disordered gas} (high $\phiA$ for $\Npc \geq 4$, or very
  low $\phiA$ for selected large $\Npc$): neither global nor local polar order, no
  enstrophy peak, Poisson-like density fluctuations.
\end{itemize}
Rather than a single window that closes with increasing $\Npc$,
active turbulence and the swarm phase interleave in $(\Npc,\phiA)$,
with turbulence reappearing at higher $\phiA$ for $\Npc = 10, 12$
after an intervening swarm window; swarm and disordered states occupy
essentially remaining space.
Phase boundaries were assigned based on snapshots, steady-state polar
order, enstrophy spectra, and velocity correlation data; the dotted
lines in Fig.~\ref{fig:phasediag}b are approximate guides and will
sharpen with longer simulations and finer parameter sweeps.
We describe each phase in turn below.

\begin{figure}
  \centering
  \includegraphics[width=0.64\columnwidth]{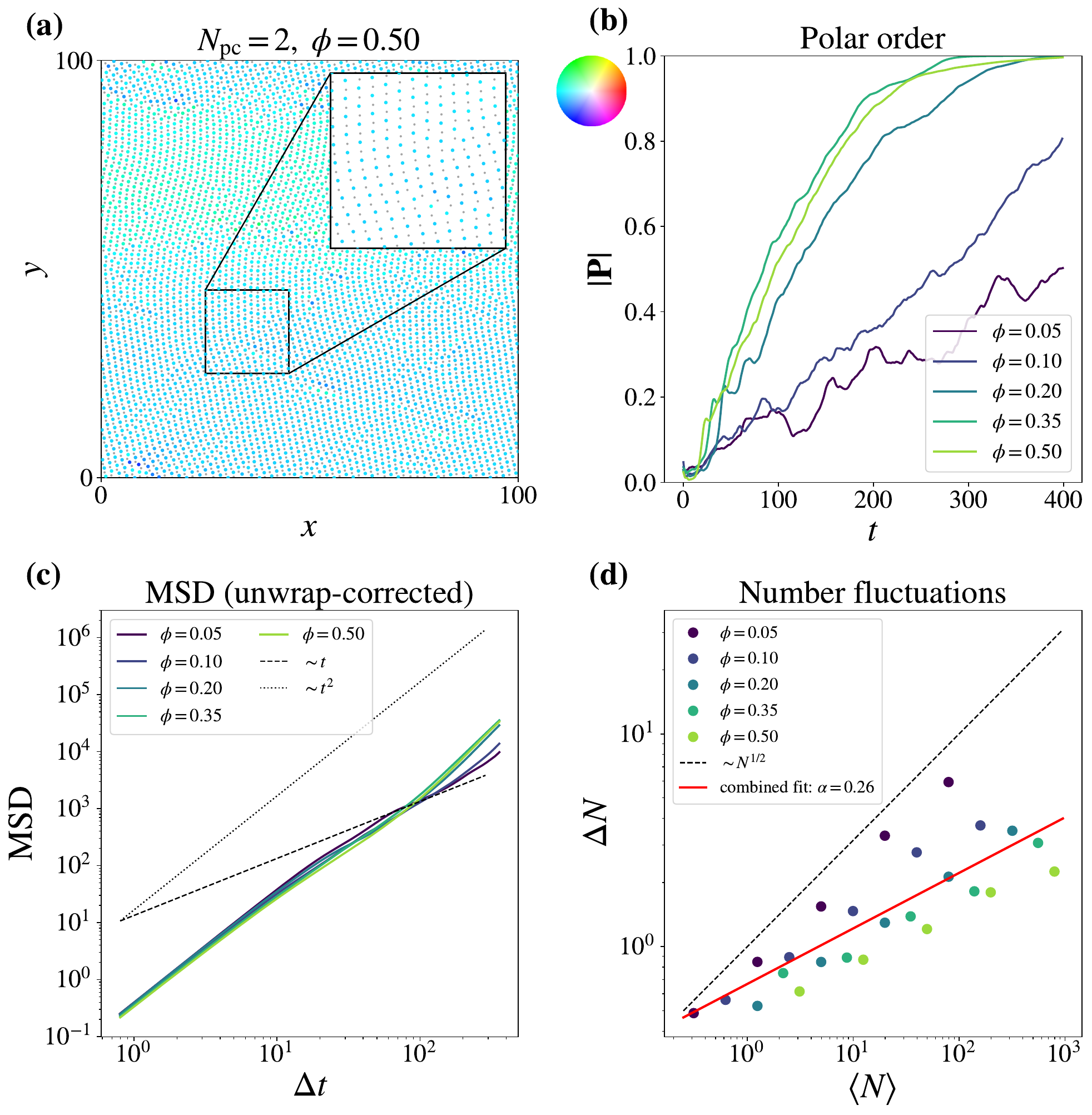}
  \caption{%
  \textbf{Polar flock ($\Npc = 2$, $J_0 > 0$).}
  (a) Steady-state snapshot at $\phi = 0.50$, $t = 0.99T$, with zoom
  inset; colour encodes head orientation $\theta\ \mathrm{mod}\ \pi$ as
  indicated by the colour wheel. The near-uniform colouring confirms
  long-range polar alignment at steady state.
  (b) Polar order $\Polv(t)$ overlaid across every density reported for
  this phase ($\phi = 0.05$--$0.50$). $\phi = 0.20$--$0.50$ approach
  $\Polv \to 1$ within the simulated window; $\phi = 0.05, 0.10$ rise
  more slowly and have not yet saturated by $t = 300$.
  (c) Mean-squared displacement of head monomers, overlaid across the same densities: the
  curves track the dotted $\sim t^2$ guide closely across the accessible
  range, confirming near-ballistic transport in the flock, consistent
  across all densities shown.
  (d) Number fluctuations $\DeltaN$ vs $\Navg$ for the same densities,
  with a single power-law fit pooling all of them: $\alpha \approx 0.30$,
  sub-Poissonian.
}
  \label{fig:flock}
\end{figure}

\subsection{Polar flocking of active dimers}
For short chains ($\Npc = 2$), the repulsive system shows a striking
dynamical behaviour (Fig.~\ref{fig:flock}).
The polar order parameter $\Polv(t)$ grows monotonically from zero
across all densities studied ($\phiA = 0.05$--$0.50$, Fig.~\ref{fig:flock}b):
at $\phiA = 0.20$--$0.50$, $\Polv$ approaches $\approx 1$ within the
simulated window, indicating the spontaneous emergence of global polar
order, while at $\phiA = 0.05, 0.10$ the growth is markedly slower and
has not yet saturated by $t = 300$.
Before reaching the flock, the system passes through a transient
disordered phase visible as a shoulder/plateau in $\Polv(t)$, where the
velocity field shows chaotic mesoscale structure qualitatively similar
to active turbulence.
The mean-squared displacement (Fig.~\ref{fig:flock}c), corrected for
periodic-boundary wrapping, is near-ballistic across the accessible time
window and consistent across all densities studied, confirming
long-range directed motion in the flock.
Number fluctuations, pooled across all densities into a single
fit, give $\alpha \approx 0.30$ (Fig.~\ref{fig:flock}d) ---
\emph{sub}-Poissonian, in contrast to the super-Poissonian exponent
predicted by Toner-Tu theory for generic two-dimensional polar active
fluids~\cite{toner1995long}.
Thus, our results indicate
that density fluctuations are strongly suppressed (the hallmark of hyperuniformity \cite{torquato2016hyperuniformity, subramaniam2026shape}).
The long-range repulsive interactions during the flocking state can lead to suppression of density fluctuations.
We return to this discrepancy and its
physical origin in Sec.~\ref{sec:theory}.
At high area fraction ($\phiA = 0.65$), crowding frustrates alignment
and polar order is destroyed, showing a re-entrant disordering as density
increases.
Polar order in point phoretic colloids via chemical repulsion has been
established~\cite{das2024flocking, subramaniam2025minimal, adhikary2025flocking}.
Steric flocking of rigid rods is well studied~\cite{Peruani2006,Ginelli2010,
grossmann2020particle}.
Our finding here is distinct from these --- polar flocking of flexible active dimers driven by phoretic
repulsion, with re-entrant disordering at
high density.

\begin{figure}
  \centering
  \includegraphics[width=0.98\columnwidth]{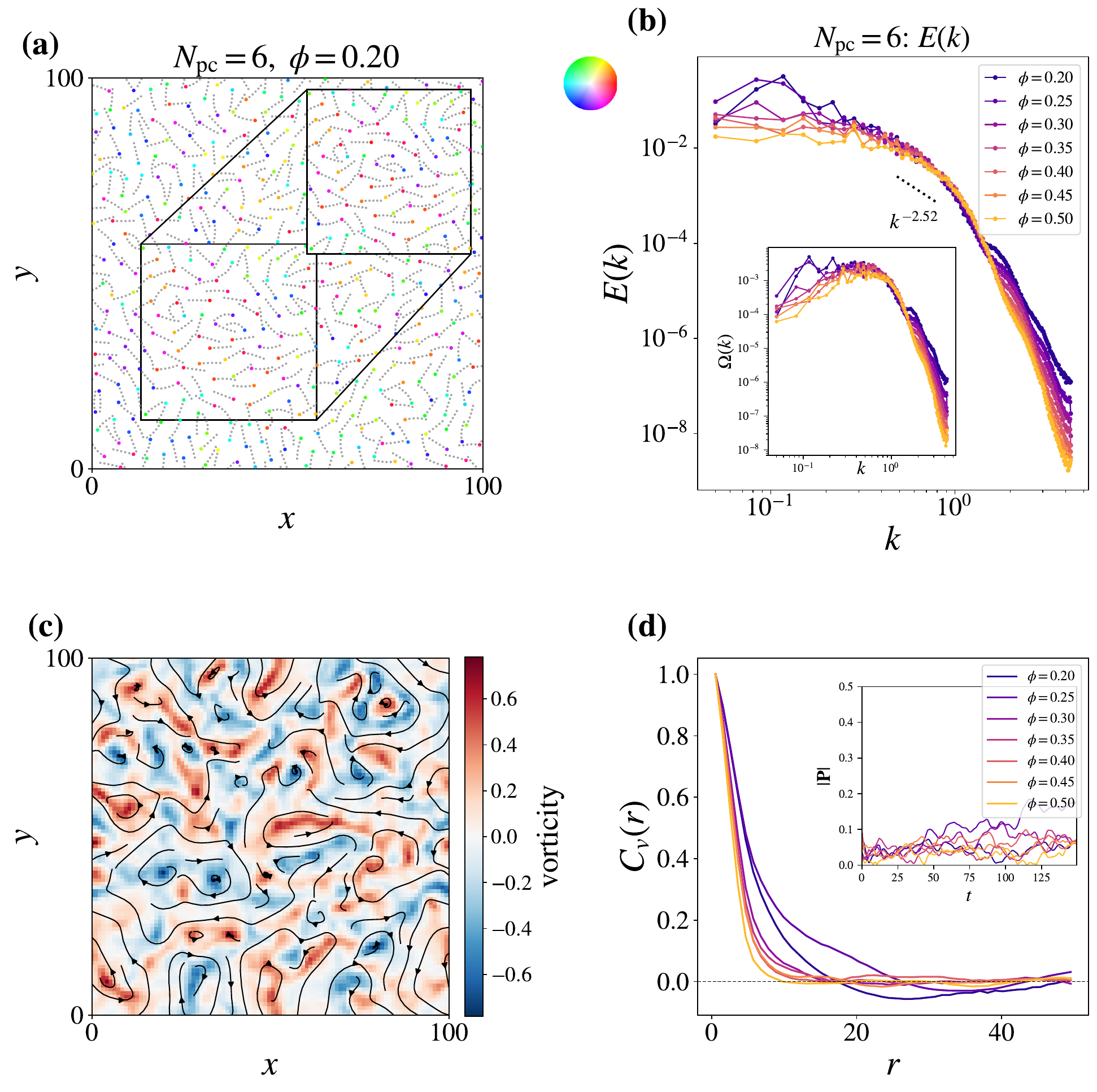}
  \caption{%
  \textbf{Active turbulence ($\Npc = 6$, $\phi = 0.20$, $J_0 > 0$).}
  (a) Steady-state snapshot with zoom inset; colour encodes head
  orientation $\theta\ \mathrm{mod}\ \pi$ as indicated by the colour
  wheel. Orientations are disordered at all scales, with no global
  alignment.
  (b) Energy spectrum $\Ek$ for every density probed at $\Npc = 6$
  ($\phi = 0.20$--$0.50$); the dotted guide line marks a fitted
  high-$k$ decay slope, shown for visual reference across the sweep. \textit{Inset}: $\Omega(k)$ spectrum.
  (c) Vorticity field (colour) with velocity streamlines overlaid,
  showing counter-rotating vortices at the mesoscale, for the
  representative point $\Npc = 6$, $\phi = 0.20$.
  (d) Velocity correlation $\Cv$ overlaid across the same $\Npc = 6$
  density sweep as (b), with polar order $\Polv(t)$ inset (truncated
  to the shortest common run duration across the sweep): the negative
  minimum present at low $\phi$ vanishes with increasing density,
  directly visualizing the turbulence-to-swarm crossover.
  Note that the ($\chi_R$, $\chi_T$) values used here are
  $\chi_R = 8$, $\chi_T = 4$.
}
  \label{fig:turbulent}
\end{figure}

\subsection{Dry route to active turbulence}

Active turbulence has been reported in rigid self-propelled rod
suspensions~\cite{Wensink2012PNAS,wensink2012emergent} and in hydrodynamically
resolved squirmer rod simulations~\cite{Zantop2022}.
In both cases the mechanism requires either alignment-disrupting
hydrodynamic torques or soft long-range Yukawa potentials at high
P\'{e}clet number.
Figure~\ref{fig:turbulent} shows that repulsively interacting chains at
intermediate aspect ratio ($\Npc = 6$, $\phiA = 0.20$) develop a chaotic
mesoscale flow state characterized by:
(a) a disordered snapshot with no large-scale alignment;
(b) an energy spectrum $\Ek$, shown here across the full $\Npc = 6$
density sweep for context, with a fitted high-$k$ decay slope shown
for visual reference;
(c) a vorticity field showing counter-rotating mesoscale vortices;
(d) a velocity correlation $\Cv$ with a clear negative minimum at
$r^* \approx 20\bead$ for the turbulent densities, vanishing at higher
$\phiA$ as the system crosses over into the swarm phase; the inset shows
polar order $\Polv$ remaining low for the turbulent densities and rising
for those in the swarm regime, confirming the absence of global alignment
specifically in the turbulent window. The negative minimum is the key diagnostic: it signals that particles
separated by $r^*$ are on average moving in \emph{opposite} directions,
as expected for particles sitting on opposite sides of a vortex pair.
The presence of counter rotating vortices, which dissipate energy at a characteristic scale, without a global polar order, is a signature of the turbulent phase ~\cite{Alert2022}. As we show in the next subsection, the presence of some sort of global polar order suppresses global vortex formation, and one instead has the \textit{swarm} phase, where local polar order exists in addition to global spatial disorder.

We note that our vortex formation mechanism is distinct from prior routes: the phoretic torque
$\chiR(\hat{\bm e} \times \mathbf{J})_z$ continuously destabilizes locally
aligned clusters by rotating heads away from approaching neighbours,
generating vorticity cascades at the cluster scale without any fluid
mediation.
This constitutes, to our knowledge, the first demonstration of active
turbulence driven purely by chemical interactions between the tips of
flexible chains in a fully dry model.

\begin{figure}
  \centering
  \includegraphics[width=0.98\columnwidth]{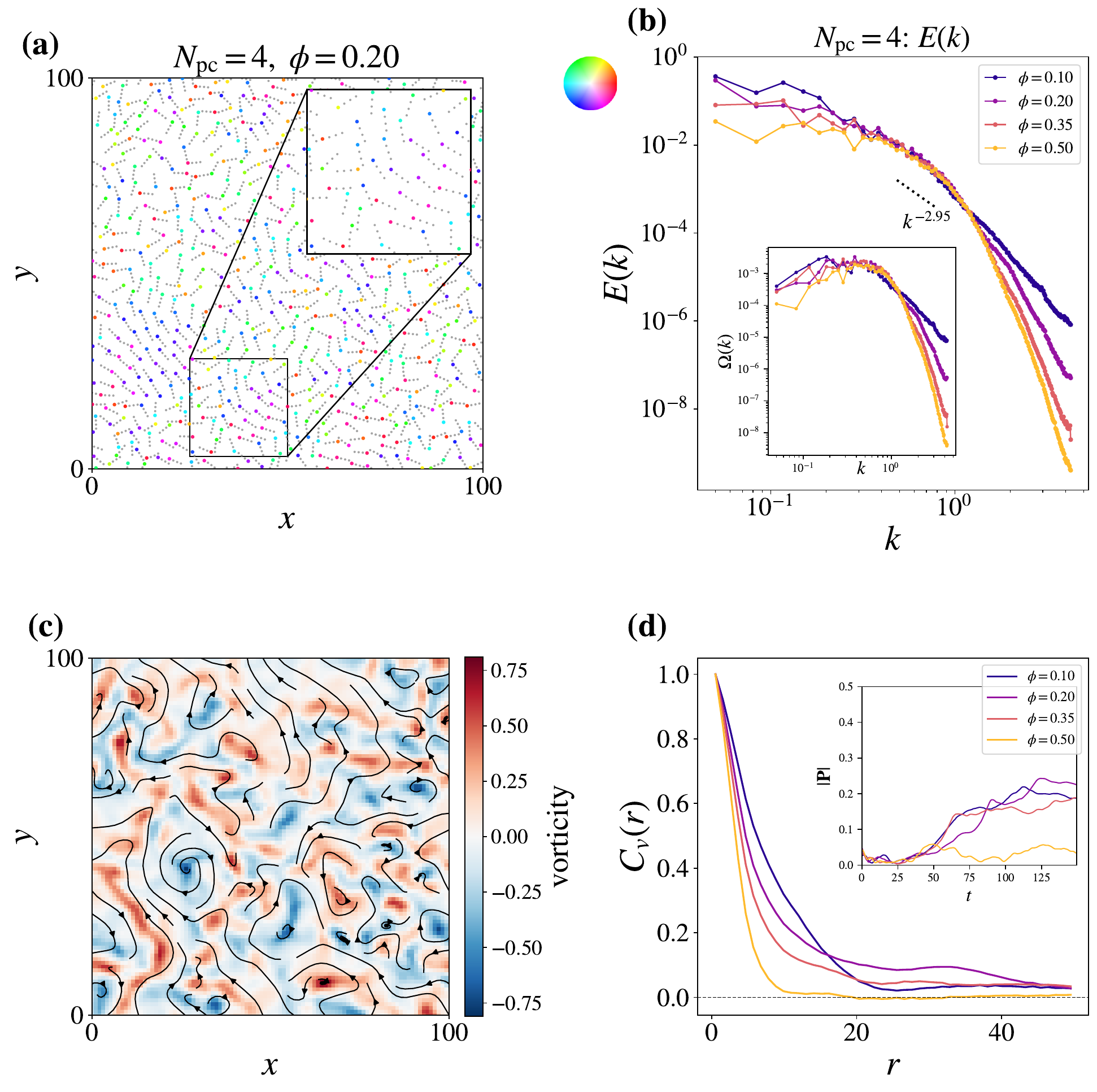}
  \caption{%
  \textbf{Swarm phase ($\Npc = 4$, $\phi = 0.20$, $J_0 > 0$).}
  (a) Steady-state snapshot with zoom inset on a region of local polar
  order; colour encodes head orientation $\theta\ \mathrm{mod}\ \pi$ as
  indicated by the colour wheel. Locally co-moving clusters are
  visible but no global polar order is established.
  (b) Energy spectrum $\Ek$ across the standard density grid at
  $\Npc = 4$ ($\phi = 0.05$--$0.65$); the dotted guide line marks a
  fitted high-$k$ decay slope, shown for visual reference across the
  sweep. \textit{Inset}: $\Omega(k)$ spectrum.
  (c) Vorticity field with velocity streamlines overlaid for the
  representative point, showing mesoscale vortices comparable in size
  to those in Fig.~\ref{fig:turbulent}c.
  (d) Velocity correlation $\Cv$ overlaid across the same density grid
  as (b), with polar order $\Polv(t)$ inset (truncated to the
  shortest common run duration): $\Cv$ decays without a negative
  minimum at every density shown, distinguishing this phase from true
  turbulence, while $\Polv$ settles at a non-zero steady-state value
  reflecting persistent local polar domains.
}
  \label{fig:swarm}
\end{figure}

\subsection{Swarming phase}

At low to intermediate densities and intermediate chain lengths
($\Npc = 4$, $\phiA = 0.20$), the system enters a qualitatively
distinct phase (Fig.~\ref{fig:swarm}).
Locally co-moving clusters are visible in the snapshot
(Fig.~\ref{fig:swarm}a).
The enstrophy spectrum (computed but not shown) peaks at the
cluster scale, comparable to the turbulent phase.
However, the velocity correlation $\Cv$ (Fig.~\ref{fig:swarm}d)
decays to zero without a negative minimum --- in contrast to the
turbulent phase --- across the whole $\Npc = 4$ density grid shown, and
the inset shows the polar order settling at a
non-zero steady-state value $\Polv \approx 0.2$--$0.3$.
This coexistence of local polar order, vortical flow, and the absence
of a $\Cv$ dip defines the swarm as a distinct phase from both
the polar flock and the turbulent state. 
Though smaller scale vortices do exist in this phase (Fig.~\ref{fig:swarm}c), the absence of negative velocity correlations is consistent with locally polar
domains suppressing the global anti-correlated flow that characterizes
pure turbulence.\\

We briefly summarize various turbulent-like
states reported in this work, which differ from one another in
physically meaningful ways.
At $\Npc = 8$, and only at $\phiA = 0.10$ and $0.20$ specifically, the
system displays conventional active turbulence signatures in steady
state: counter-rotating vortices visible in the velocity field, an
enstrophy peak at $k^* \approx 0.3\bead^{-1}$ (computed but not shown),
and a negative minimum in the velocity correlation $\Cv$ at
$r^* \approx 20\bead$.
Every other density and chain length we tested in this range --- including
the remaining $\Npc = 8$ densities and all of $\Npc = 10, 12$ --- shows the
swarm signature instead (no $\Cv$ minimum); active turbulence is thus a
narrow window that closes with increasing chain length, rather than a
broad intermediate-$\Npc$ regime.
The swarm ($\Npc = 4$, low $\phiA$) shares the vortical flow
and enstrophy peak with the turbulent phase, but the $\Cv$ dip is
absent - due to the aforementioned co-existence between local polar order and local counter-vortices.
This distinction between locally-polar turbulent flow and globally-disordered
turbulent flow is a direct consequence of the chain geometry: shorter
chains have a larger head-to-body ratio and a stronger effective
self-propulsion per unit drag, making local alignment easier to sustain
even in the presence of chemical reorientation torques.
Though not studied in more detail here, we also note that active dimers ($\Npc = 2$) display \emph{transient} turbulence: the
chaotic mesoscale flow is not a steady-state but a kinetic precursor
to global polar order; it disappears once alignment is established.

\section{Attractive phoretic interactions}\label{sec:attractive}
In this section, we consider the emergent dynamics of colloidal chains with active-tips that mediate attractive phoretic interactions ( $J_0 < 0$). 
Results are summarized in Fig.~\ref{fig:phasediag}(c), (g)--(i).
The three regimes are identified for $J_0 < 0$ are summarize below:
\begin{itemize}
  \item \emph{Micellar phase} ($\phiA \lesssim 0.20$, most chain
  lengths): isolated hedgehog-like aggregates with heads inward and
  tails outward, stabilized by attractive phoretic interactions.

  \item \emph{Liposomal/stacked} ($\Npc = 2$, $\phiA \lesssim 0.20$): a
  layered variant specific to the shortest chains, in which micellar
  aggregates organize into liposomal-like concentric structures.

  \item \emph{Active glass} ($\phiA \gtrsim 0.35$, all chain lengths): a
  dense, dynamically-arrested, locally-ordered packing; the percolation and pair-correlation
  diagnostics in Fig.~\ref{fig:attractive}d--f show these system is structurally a solid, with slow dynamical rearrangement.
\end{itemize}
When $\Jzero < 0$, phoretic interactions are attractive: heads are
pulled toward each other and are rotationally reoriented to point inward.
In contrast to the repulsive case, the dynamics are aggregative:
heads cluster, driven by the chemical gradient, while tails are dragged
along as passive appendages. These are distinguished structurally and dynamically in
Fig.~\ref{fig:attractive}: by aggregate morphology in the snapshots
(a)--(c); by percolation behaviour (d), which separates isolated finite
aggregates (micellar, liposomal) from a single space-spanning network
(active glass); and by pair-correlation structure (e,f), which tests
directly for the local hexatic order suggested by the active-glass
snapshots.
Phase boundaries were assigned based on snapshots, steady-state polar
order, and structural/percolation diagnostics.
We describe each phase in turn below.

\subsection{Micellar and liposomal phases}

For $\Npc = 6$ at low density ($\phiA = 0.10$, Fig.~\ref{fig:attractive}a),
compact hedgehog-like aggregates form in which heads occupy a dense core and
tails radiate outward --- a spatial arrangement directly analogous to an
amphiphile micelle, with the chemically active head playing the role of
the polar headgroup and the flexible tail that of the hydrophobic chain.
The assembly is entirely non-equilibrium: it requires the active
self-propulsion at the head to sustain the inward orientation.
The zoom inset of Fig.~\ref{fig:attractive}a reveals the internal
radial structure clearly, with heads concentrated at the center and
tails fanning outward.

For $\Npc = 2$ at $\phiA = 0.20$ (Fig.~\ref{fig:attractive}b), a
qualitatively different \emph{stacked} or liposomal phase emerges, in
which micellar aggregates organize into layered arrangements reminiscent
of a two-dimensional multilamellar or liposomal phase.
The zoom inset shows the internal concentric ring structure of a single
aggregate, with heads forming the innermost core and successive rings
of tail monomers radiating outward.
The emergent amphiphilic geometry of the dimer --- head-attracted,
tail-excluded --- produces this hierarchy of mesoscale structures from
a purely phoretic drive. These first two phases, besides being visually distinct, are distinguished by studying the clustering dynamics of the tip monomers (see Appendix \ref{app:methods:clustering}). As shown in Fig. \ref{fig:attractive}(d), the micellar phase has a large number of clusters, though the largest cluster size is relatively small, whereas the liposomal phase has a small number of clusters with the largest cluster size (the outer concentric ring) the largest compared to all other phases. The structure of the micellar aggregates in (a) is also well captured by the angular pair correlation (Fig. \ref{fig:attractive}(e) - inset) and structure factor (Fig. \ref{fig:attractive}(f) - inset), where the peak in the former indicates a long-ranged repetitive order of this phase.

The structural analogy with equilibrium amphiphile self-assembly is
non-trivial.
Block copolymer micelles~\cite{Laradji2004} and Janus nanoparticles
adsorbed on lipid vesicles~\cite{Zhu2023SoftMatter} organize into
Platonic-solid and lamellar geometries via curvature-mediated or
hydrophobic/hydrophilic interactions; our active system reaches
analogous structures via a non-equilibrium chemical drive.
Amphiphatic colloidal spheres have been shown to self-assemble into
hollow faceted cages~\cite{Miller2009}; active amphiphilic Janus
particles produce analogous self-assembly modulated by
activity~\cite{Mallory2017NJP}.
Our system differs from all of these in that the amphiphilic character
is \emph{emergent} rather than designed: a symmetric chain with a single
active tip spontaneously acquires head-tail asymmetry through the
phoretic interaction, with no built-in hydrophobic or geometric
asymmetry between head and tail monomers.\\

\subsection{Active glass phase}

\begin{figure}
  \centering
  \includegraphics[width=0.98\columnwidth]{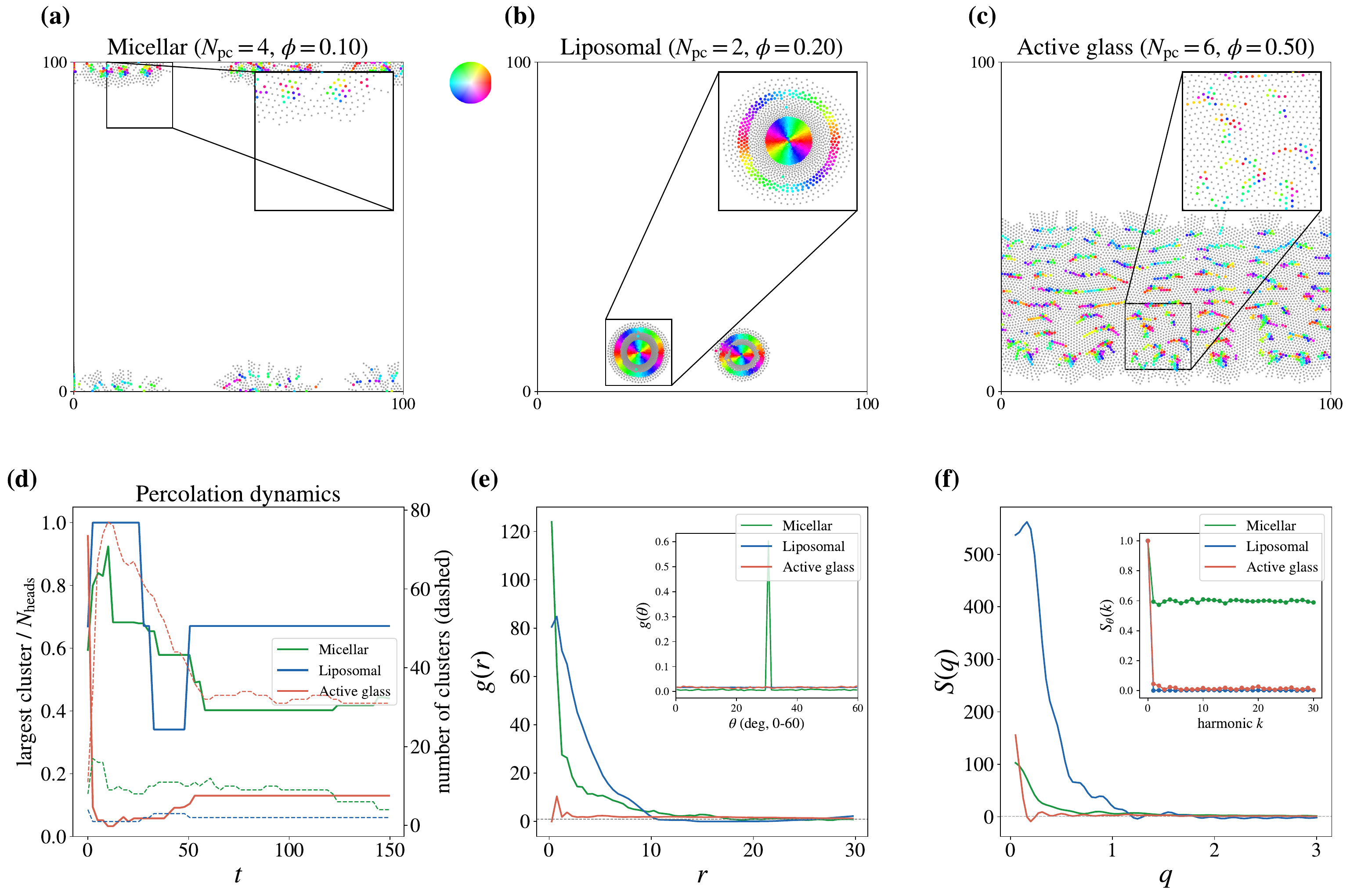}
  \caption{%
  \textbf{Attractive phoretic interactions ($J_0 < 0$): phases and structural
  diagnostics.}
  (a)--(c) Steady-state snapshots with zoom insets: micellar
  ($\Npc = 4$, $\phi = 0.10$), hedgehog-like aggregates with heads at
  the core and tails radiating outward; liposomal/stacked
  ($\Npc = 2$, $\phi = 0.20$), concentric ring structure characteristic
  of an active non-equilibrium micelle; active glass ($\Npc = 6$,
  $\phi = 0.50$), a dense, space-filling packing.
  (d) Percolation dynamics: fraction of heads in the largest connected
  cluster $f_{\max}(t)$ (solid) and number of distinct clusters (dashed,
  right axis), overlaid across all three phases. Micellar and liposomal
  aggregates plateau at a small, stable $f_{\max}$ (isolated, finite
  aggregates); active glass shows a persistently large/growing
  $f_{\max}$, consistent with a single space-spanning network rather
  than isolated clusters --- the diagnostic that actually distinguishes
  these phases, since the local clustering coefficient alone cannot.
  (e) Radial pair correlation $g(r)$, with the angular bond distribution
  $g(\theta)$ (folded into $[0^\circ,60^\circ)$) inset: active glass
  shows structure extending to multiple neighbour shells in $g(r)$ and
  a sharp peak in $g(\theta)$, the direct real-space signature of local
  hexatic order, absent in the isolated micellar/liposomal aggregates.
  (f) $S(q)$, the Hankel transform of the $g(r)$ in (e), with the
  Fourier spectrum of $g(\theta)$ inset (harmonic index $k$): the
  reciprocal-space counterpart of (e), confirming the same structural
  distinction.
}
  \label{fig:attractive}
\end{figure}

At higher density (Fig.~\ref{fig:attractive}c,d), the attractive system
produces progressively more arrested configurations.
At $\phiA = 0.50$ (Fig.~\ref{fig:attractive}c), chains aggregate into
dense disordered clusters with short-range nematic order; the zoom inset
reveals that chains within each cluster are locally aligned, with heads
pointing toward a common center.
At $\phiA = 0.65$ (Fig.~\ref{fig:attractive}d), a kinetically arrested
quasi-lamellar state forms whose structure depends on initial conditions,
suggestive of a non-equilibrium glass or gel. It is evident from both the snaphots and the pair correlation function (Fig. \ref{fig:attractive}(e) - red) that the spatial structure of this phase is effectively that of a crystal, though the dynamics is strongly arrested.
Though highly arrested at the time scale of simulations, these states require substantially longer simulations
to determine whether they represent true dynamical steady states. We note that direct experimental realization of the attractive case is
more challenging than the repulsive one: the
system in ~\cite{Shelke2026Science} is repulsive (pusher stresslet at the head).
Realizing attractive tip phoresis would require, for instance, a head that
consumes rather than produces fuel, tuning the relevant chemical mobilities ~\cite{meredith2020predator}.

\section{Coarse-grained description}
\label{sec:theory}

\subsection{Kinetic theory framework}

Several theoretical frameworks have been developed to derive hydrodynamic
equations for collections of self-propelled elongated bodies from
microscopic equations of motion, broadly falling into two strands.
One strand derives continuum equations via a one-body kinetic equation
for the distribution function $\psi(\bm{R},\theta,t)$, with active
currents specified directly from the microscopic pair
interaction~\cite{LM2003,LM2005}; this approach is particularly suited
to systems where the interaction is a specified function of the
relative orientation and position of two colloidal chains, and has also been
applied to self-propelled hard rods, deriving a Boltzmann-like kinetic
equation and showing that steric collisions alone produce nematic but
not polar order~\cite{BM2008,BM2008PRL}.
A complementary technique is modelling active particles directly as
continuum fields, capturing the transition between disordered and
polar-ordered phases for self-propelled ellipsoids via an effective
shape parameter (eccentricity) that enters the collisional
cross-section~\cite{grossmann2020particle}.

Our system differs from both of these in two essential ways.
First, the interaction is \emph{chemical} (monopolar phoretic field
from the tip), not steric or hydrodynamic.
Second, the active body is a \emph{flexible} bead-spring chain rather
than a rigid ellipsoid or filament.
We thus adopt the rigid-rod model of \cite{LM2003, BM2008PRL}, and extend it to our system with chemically emitting tips.
We note that this is a simplification of our semi-flexible chains - here, each chain of $\Npc$
monomers is replaced by a rigid polar rod of length
$\ell = (\Npc-1) \times 2\bead$ and orientation $\theta$. The consequences of this simplification will be discussed below.
The active tip (head monomer) is located at
$\bm{r}_{\rm tip} = \bm{R} + \tfrac{\ell}{2}\,\nhat$,
where $\bm{R}$ is the rod center of mass: the head sits at one end of
the chain, and for a uniform rigid rod the center of mass is at the
geometric center, $\ell/2$ from either end.
The tip offset therefore grows with chain length, and $\ell$ enters
the theory both through this offset and through the hard-rod
excluded-volume domain.
The full derivation is given in Appendix~\ref{app:theory}; we
summarise the key steps and results here.


\subsection{One-body kinetic equation and moment closure}

The one-body distribution $\psi(\bm{R},\theta,t)$ obeys a conservation
law whose rotational current receives two contributions: rotational
diffusion and the phoretic pair torque.
In the mean-field (molecular chaos) approximation the pair distribution
factorises as $f_2 = \psi(\bm{R},\theta)\psi(\bm{R}',\theta')$, and the
phoretic rotational current becomes
\begin{equation}\label{eq:Jrph_main}
  J^{\theta,\mathrm ph}(\bm{R},\theta)
  = \int_{\mathcal{D}(\theta,\theta')} d^2\xivec \int_0^{2\pi} d\theta'\;
    \Omega(\xivec,\theta{-}\theta';\ell)\;
    \psi(\bm{R},\theta)\,\psi(\bm{R}-\xivec,\theta'),
\end{equation}
where $\xivec = \bm{R} - \bm{R}'$ is the center-to-center separation and
the domain $\mathcal{D}$ enforces the hard-rod excluded-volume condition
\begin{equation}\label{eq:dmin_main}
  d_{\min}(\xivec,\theta,\theta';\ell)
  = \min_{s,s' \in [-\ell/2,\ell/2]}
    |\xivec + s\nhat - s'\nhat'| \geq 2\bead.
\end{equation}
This is the correct hard-rod body-to-body distance, distinct from the
tip-to-tip distance.
The pair torque is
\begin{equation}\label{eq:Omega_main}
  \Omega(\xivec,\Delta\theta;\ell)
  = \chiR\Jzero\,
    \frac{(\nhat \times \xivec)_z + \bead\sin\Delta\theta}
         {|\xivec + \frac{\ell}{2} (\nhat - \nhat')|^3},
  \quad \Delta\theta = \theta - \theta'.
\end{equation}

Taking the first angular moment of the kinetic equation and retaining
only the $k=0$ (spatially uniform) mode, the polarization 
$\bm{P}$
equation
reduces to
\begin{equation}\label{eq:Pgrowth_main}
  \partial_t \bm{P}
  = \left[-\frac{\rho_{\mathrm rod} K_1(\ell)}{2} - \Dr\right]\bm{P} + \ldots,
\end{equation}
where $\rho_{\mathrm rod} = \phi/(N_{\mathrm pc}\pi\bead^2)$ is the rod number
density and $K_1(\ell)$ is the \emph{polar alignment coefficient},
the $\sin\Delta\theta$ Fourier mode of the position-averaged pair torque:
\begin{equation}\label{eq:K1_main}
  K_1(\ell)
  = \frac{1}{\pi}\int_0^{2\pi} d(\Delta\theta)\;\sin(\Delta\theta)
    \int_{\mathcal{D}(\Delta\theta)} d^2\xivec\;
    \Omega(\xivec,\Delta\theta;\ell).
\end{equation}
Here $\mathcal{D}(\Delta\theta)$ is the hard-rod excluded-volume domain
(Eq.~\eqref{eq:dmin_main}); at the linear, homogeneous level considered
here it enters $K_1(\ell)$ only through this domain restriction, since
the Onsager steric torque itself is apolar and contributes zero polar
Fourier coefficient~\cite{Onsager}, consistent with steric-only
mechanisms producing nematic rather than polar
order~\cite{BM2008,BM2008PRL}.

\subsection{Physical interpretation of $K_1$}

The sign of $K_1$ determines the nature of the phoretic torque.
For $K_1 < 0$, the growth rate in Eq.~\eqref{eq:Pgrowth_main} is
positive at sufficient density: the phoretic interaction is
\emph{polar-aligning}.
For $K_1 > 0$, the growth rate is always negative: the torque is
\emph{anti-aligning} and no polar instability exists.

The physical origin of the sign is geometric.
For repulsive phoresis ($\Jzero > 0$), the concentration gradient at
the tip of rod $i$ points \emph{away} from rod $j$'s tip.
When two nearly-parallel rods approach, this gradient pushes rod $i$'s
tip away from rod $j$'s tip, rotating rod $i$ back toward alignment with
rod $j$ --- a polar-aligning torque, giving $K_1 < 0$.
For attractive phoretic interactions ($\Jzero < 0$), the gradient pulls the tip
toward rod $j$, rotating rod $i$ \emph{away} from rod $j$'s orientation
--- anti-aligning, giving $K_1 > 0$.

Crucially, $K_1(0) = 0$ for point particles ($\ell = 0$): the tip offset
$\frac{\ell}{2}$ is what breaks the orientational symmetry that makes point monopoles
non-aligning.
The polar instability is therefore a \emph{geometric} effect, arising
from the spatial asymmetry between the chemical source position and the
rod center of mass.

\begin{figure}
  \centering
  \includegraphics[width=\columnwidth]{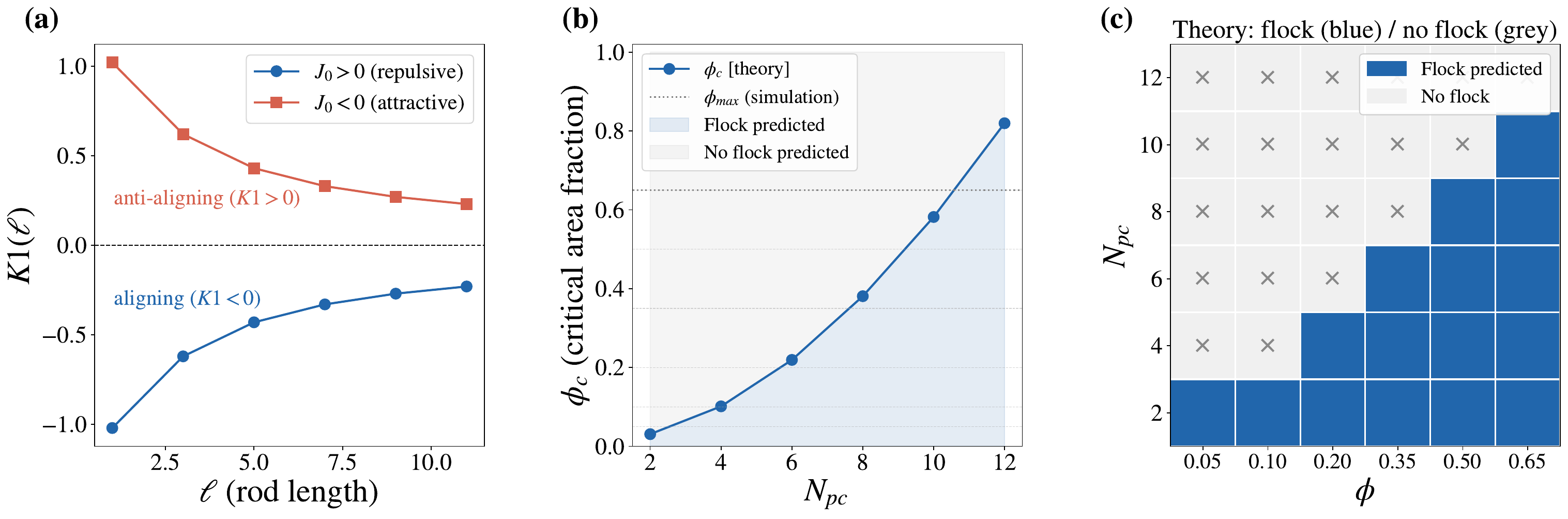}
  \caption{%
    \textbf{Rigid-rod kinetic theory of tip-active phoretic chains.}
    (a) Polar alignment coefficient $K_1(\ell)$ vs rod length $\ell$
    (equivalently, $\Npc$) for repulsive ($\Jzero > 0$, blue circles) and
    attractive ($\Jzero < 0$, red squares) phoresis, computed numerically
    from Eq.~\eqref{eq:K1_main} with the correct hard-rod excluded-volume
    domain (Eq.~\ref{eq:dmin_main}).
    For repulsive phoresis $K_1 < 0$ (polar-aligning) for all $\Npc$,
    with magnitude decreasing with chain length.
    For attractive phoretic interactions $K_1 > 0$ (anti-aligning) for all $\Npc$,
    predicting no polar instability at any density.
    (b) Critical area fraction $\phi_c(\Npc) = 2\pi\bead^2\Npc\Dr/|K_1(\ell)|$
    for the onset of polar flocking (repulsive case) vs $\Npc$.
    The dotted horizontal line marks the maximum area fraction explored
    in simulation ($\phi = 0.65$); the shaded region below the curve
    is where flocking is predicted.
    $\phi_c$ increases steeply with $\Npc$, reflecting the decrease of
    $|K_1|$ with chain length.
    (c) Binary matrix in $(\phi, \Npc)$ space showing where the theory
    predicts polar flocking ($\phi > \phi_c(\Npc)$, blue, 
    vs no flocking ($\phi < \phi_c(\Npc)$, grey. 
    The theory correctly predicts flocking only for short chains at
    moderate-to-high density ($\Npc = 2$ at all simulated $\phi$;
    $\Npc = 4$ at $\phi \gtrsim 0.35$) and no flocking for long chains
    ($\Npc \geq 10$ at all simulated $\phi$), in qualitative agreement
    with the simulation phase diagram (Fig.~\ref{fig:phasediag}b).
  \label{fig:theory}
  }
\end{figure}

\subsection{Comparison with simulation}

Figure~\ref{fig:theory} shows $K_1(\ell)$ (panel a) and the critical
area fraction $\phi_c(\Npc)$ (panel b) computed numerically from
Eq.~\eqref{eq:K1_main}. For repulsive phoresis, $K_1 < 0$ for all $\Npc$, confirming polar
alignment at all chain lengths within the dilute mean-field theory.
The magnitude $|K_1|$ decreases monotonically with $\Npc$, reflecting
the increasing excluded-volume domain that suppresses near-contact
torque configurations for longer chains.

The critical area fraction
$\phi_c(\Npc) = 2\pi\bead^2\Npc\Dr/|K_1(\ell)|$
increases steeply with $\Npc$ (Fig.~\ref{fig:theory}b):
$\phi_c \approx 0.04$ for $\Npc = 2$ and $\phi_c \approx 0.87$ for
$\Npc = 12$.
Since the simulation explores $\phi \leq 0.65$, the theory predicts:
\begin{itemize}
  \item $\Npc = 2$: $\phi_c \ll \phi_{\mathrm sim}$ --- flocking expected
        at all simulated densities, consistent with the observed global
        polar flock (Sec.~\ref{sec:repulsive}).
  \item $\Npc = 4$--$6$: $\phi_c$ is within the simulation range ---
        flocking is marginal and density-dependent, consistent with the
        swarm phase at moderate density.
  \item $\Npc \geq 10$: $\phi_c > \phi_{\mathrm max}$ --- no polar instability
        at any simulated density, consistent with the observed turbulent
        and disordered phases.
\end{itemize}

The rigid-rod kinetic theory thus correctly captures the qualitative
$\Npc$-dependence of the polar instability threshold.
The rigid-rod approximation is exact for $\Npc = 2$, since the dimer
and rigid rod limits coincide for a stiff two-bead chain; the good
agreement there validates the kinetic theory framework.
For larger $\Npc$, the theory overestimates $\phi_c$ relative
to simulation, which we attribute to two effects beyond the present theory. Primarily, for longer chains, the internal bending degrees of freedom introduce
additional relaxation timescales not captured by the rigid-rod description,
effectively increasing the orientational damping and suppressing global
alignment. In addition, at the moderate-to-high densities where flocking occurs ($\phi \sim 0.3$--$0.5$),
the molecular chaos approximation fails and finite-density pair correlations
$g(\xivec,\Delta\theta;\rho)$ renormalize $K_1^{\mathrm eff}(\rho,\ell)$
away from its dilute value.

For attractive phoretic interactions ($\Jzero < 0$), $K_1 > 0$ for all $\Npc$:
the polar growth rate is always negative and no flocking instability
exists at any density.
This is immediately consistent with the simulation, where the attractive
case produces micellar aggregation rather than polar order
(Sec.~\ref{sec:attractive}).
The instability in the attractive case is in the \emph{density} channel
rather than the polar channel --- the attractive chemical gradient drives
density fluctuations to grow, producing clusters and micelles rather than
alignment. We also note that a heuristic vortex-size scaling can be obtained by
extending the same linear stability analysis to finite $k$,
$\ell_v^{\mathrm th} \sim (\rho_{\mathrm rod}|K_1(\ell)|/2D_t)^{-1/2}$,
which agrees qualitatively with the monotone decrease of the vortex
scale with $\phi$ observed in simulation (not shown here); a genuine
finite-$k$ instability would require a higher-order ($k^2$--$k^4$) term
in the gradient expansion, which we do not attempt.
Similarly, the phoretic contribution to the effective density
diffusivity, $D_{\mathrm eff} = D_t + D_{\mathrm ph}$, offers a
plausible mechanism for the sub-Poissonian ($\alpha < 0.5$) number
fluctuations observed in the polar flock, though a finite $D_{\mathrm
eff}$ gives $S(k\to0) \to \mathrm{const.}$ rather than $S(k\to0)\to0$,
so this does not establish hyperuniformity in the strict sense without
a fuller density-fluctuation theory.

\section{Summary and discussions}\label{sec:discussion}

To summarize, we have shown that flexible bead-spring chains with a single chemically
active tip, interacting via a monopolar phoretic field and with no
hydrodynamic interactions, generate a rich swathe of non-equilibrium phases, spanning polar
flocking, active turbulence, swarming, non-equilibrium micellar
assembly, and kinetically arrested dense phases. It is of importance to note that, the asymmetry of activity at the individual constituent level is required for the non-equilibrium phases that we report; for chains that are active throughout their body axis \cite{kumar2024emergent} no such phases are seen (apart from trivial ones; not shown here).

Active turbulence in many-body particulate systems has previously been attributed
to hydrodynamic torques from pusher flow fields~\cite{Zantop2022}, soft
long-range Yukawa potentials in noise-free rod suspensions~\cite{Wensink2012PNAS}, the simultaneous action of chemical and hydrodynamic fields for point
phoretic particles~\cite{Chardac2024NatComm}, or via competition between effects of crowding and persistent propulsion~\cite{Keta2024PRL}.
Our route is distinct from all of these: the turbulence emerges at
moderate densities from the phoretic torque
$\chiR(\hat{\bm e} \times \mathbf{J})_z$ acting on chain heads, which
continuously destabilizes locally aligned clusters and generates vorticity
cascades at the cluster scale without any fluid mediation or density
crowding.
This thus constitutes a novel demonstration of active turbulence driven purely by long-range chemical interactions between the tips of
flexible chains.

The swarm phase identified here is characterized by vortical flow
and an enstrophy peak co-existing with sustained local polar order and
no negative minimum in $\Cv$.
It is related to the ``polar liquid'' or ``coherent flow'' phases
reported in continuum theories of active polar
fluids~\cite{Wensink2012PNAS}, but arises here from a purely phoretic
mechanism in a discrete chain model.
The physical picture is that locally polar domains contribute a positive
velocity correlation at intermediate separations that cancels the negative
vortex contribution, producing a net $\Cv$ that is everywhere non-negative
despite the clear vortical structure in the flow field.
This phase may be relevant for experimental systems where chemical
signalling produces local alignment without global polar order, such as
dense suspensions of bacteria producing attractant gradients.

The polar flock reported here for $\Npc = 2$ --- global polar order
emerging from purely phoretic repulsion, with a turbulent kinetic
precursor --- constitutes a minimal model for flocking without alignment
rules in an extended body.
A direct experimental realization may be possible using asymmetric
colloidal dimers driven by non-reciprocal chemical exchange: the authors of ~\cite{meredith2020predator} demonstrated that asymmetric oil
droplet dimers driven by differential solubility produce directed
persistent motion with a well-defined head and tail, matching very closely our active dimer setup.
A dense suspension of such dimers in a fuel-containing medium would
realize the phoretically repulsive dimer system studied here.
Alternatively, the hematite-TPM dimers of ~\cite{palacci2013living}, where a photocatalytic hematite lobe drives
the dimer forward while a passive TPM lobe trails behind, provide a
direct experimental analogue with tip-localized chemical activity.

The emergent amphiphilic character of tip-active chains produces
micellar, liposomal, and lamellar-like structures via purely
non-equilibrium driving, without designed hydrophobic/hydrophilic
asymmetry. This is a truly non-equilibrium micellar assembly.
While direct experimental realization of attractive tip phoresis is
more challenging --- the structural analogies with equilibrium block-copolymer
micelles~\cite{Laradji2004} and Janus nanoparticle assemblies on lipid
vesicles~\cite{Zhu2023SoftMatter} suggest that the phase diagram of
tip-active chains under attractive coupling is a rich target for future
theoretical and computational study.

The rigid-rod kinetic theory of Sec.~\ref{sec:theory} captures the
qualitative $\Npc$-dependence of the phase diagram remarkably well given
its simplicity: $K_1(\ell) < 0$ for all $\Npc$ correctly predicts a
polar-aligning torque at every chain length probed, $\phi_c(\Npc)$
increasing steeply with $\Npc$ correctly predicts that only short chains
can flock within the simulated density range, and the prediction of no
polar instability at all for $\Npc \geq 10$ matches the observed absence
of any global or local polar order channel at those chain lengths
(Fig.~\ref{fig:phasediag}b). Nevertheless, this theory cannot explain all our results. The theory
predicts a polar instability at high density for every chain length
studied ($\phi > \phi_c(\Npc)$), yet no such high density flocking is
observed in simulation --- instead, these points fall in the swarm
phase (Sec.~\ref{sec:repulsive}), with sustained \emph{local} polar
order ($\Polv \approx 0.2$--$0.3$) rather than global alignment
($\Polv \to 1$).
We attribute this to the linear stability analysis only establishing
that the uniform ($k=0$) state is unstable to \emph{some} polar order,
not whether that order saturates into one system-spanning domain or
many small ones --- a nonlinear question outside its scope.
The finite domain size is therefore likely set by physics not included
here, most plausibly the excluded-volume interaction of a genuinely
semi-flexible chain rather than an idealized rigid rod, which we do
not pursue in this work.

Our model is two-dimensional and noiseless ($D_r = 0$ in most runs);
adding rotational diffusion would broaden phase boundaries and may
destroy the long-time flock (though long-range phoretic interactions
may stabilize polar order~\cite{subramaniam2025minimal}).
The chemical loop scales as $O(N_c^2)$ and limits accessible system
sizes; a multipole expansion or fast-summation scheme would enable
larger simulations approaching the thermodynamic limit.
Extending to dipolar chemical interactions (the next multipole term in
the expansion of the tip field) is a natural theoretical next step;
this would introduce a $1/r^3$ interaction and may enrich the phase
diagram further.
Finally, direct experimental realization of the attractive case is more
challenging: the system studied in~\cite{Shelke2026Science} is
repulsive (pusher stresslet at the head). Studying the competing roles of 
phoretic and hydrodynamic interactions 
\cite{keaverny2024, canio2017, laskar2017, krishnamurthy2023} 
in chains with active-tip 
suggests another avenue for future work.


\begin{acknowledgments}
A.G.S.\ acknowledges funding from the DIA Fellowship from the Government
of India.
The authors thank P.~B.~Sunil Kumar, Manoj Kumar, and Shashi Thutupalli
for discussions.
\end{acknowledgments}

\section*{DATA AVAILABILITY}
The data that support the findings of this article are generated from computer simulation.
The data are available from the authors
upon reasonable request.
All parameters used and simulation method are provided in the article. 
\appendix

\section{Time evolution of identified phases}\label{app:evolution}

\begin{figure}
  \centering
  \includegraphics[width=\textwidth]{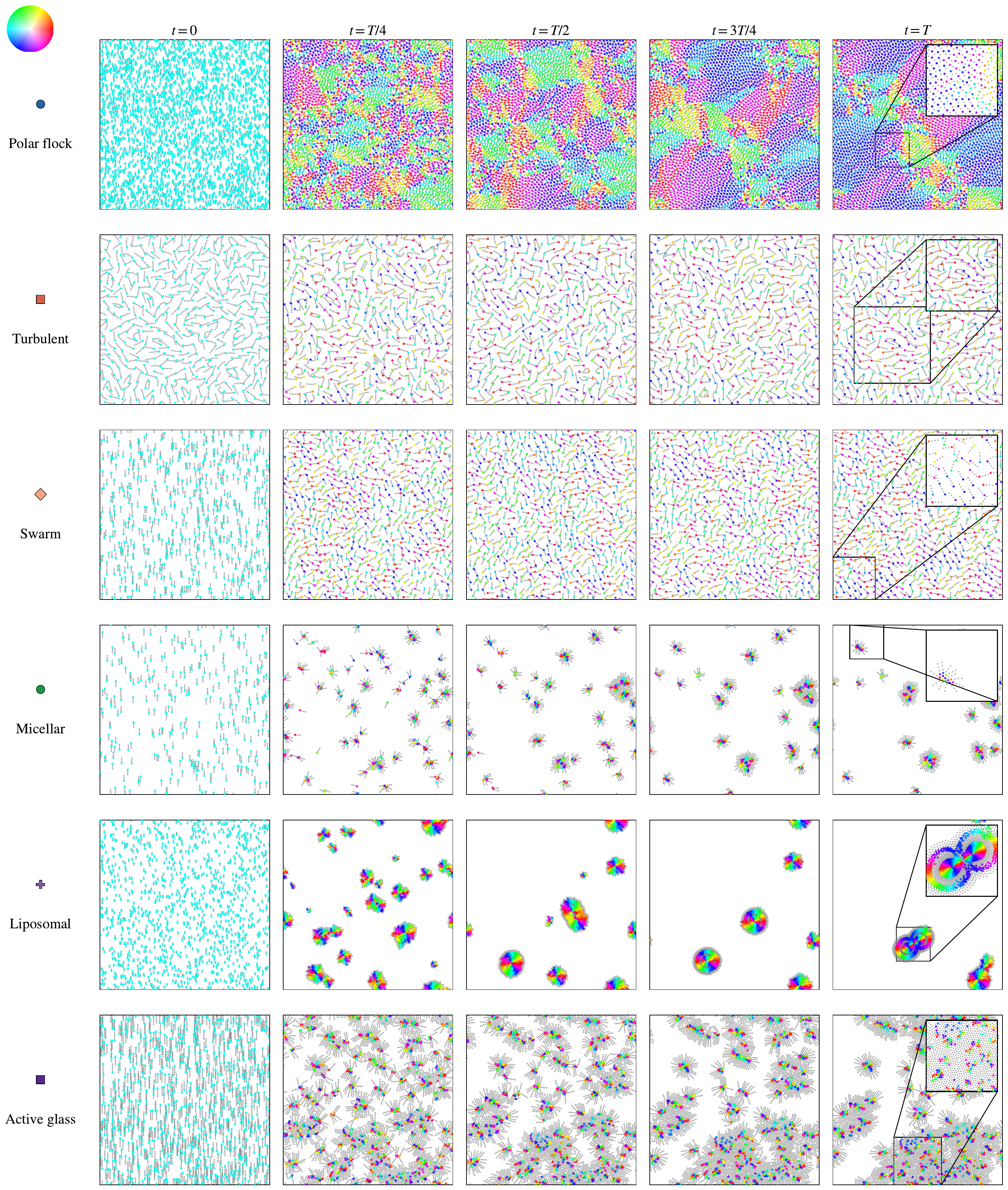}
  \caption{%
    \textbf{Time evolution of all six identified phases.}
    Each row shows one phase (marker matches the corresponding symbol in
    Fig.~\ref{fig:phasediag}b,c: polar flock, active turbulence, swarm,
    micellar, liposomal, active glass; the disordered gas is omitted), at
    representative points $\Npc, \phi$ corresponding to the snapshots in
    Fig.~\ref{fig:phasediag}d--i. Columns show the state at
    $t = 0, T/4, T/2, 3T/4, T$, where $T$ is the total simulated duration
    for that trajectory (varies by phase; see main text for individual
    values). The final column shows the full state at $t=T$ with a zoom
    inset on the same region highlighted in Fig.~\ref{fig:phasediag}.
    Colour encodes head orientation $\theta\ \mathrm{mod}\ \pi$ as
    indicated by the colour wheel.
  }
  \label{fig:appendix_evolution}
\end{figure}

Figure~\ref{fig:appendix_evolution} shows the time evolution of all six
identified phases (the disordered gas is omitted), from initialisation
to the final simulated state, at the representative $(\Npc,\phi)$ points
used for the corresponding panels of Fig.~\ref{fig:phasediag}.
This complements the steady-state snapshots in the main text by showing
how each phase is actually reached: the polar flock nucleates from an
initially disordered configuration and coarsens into a single aligned
domain; the turbulent and swarm states remain visually stationary in
their qualitative character throughout, consistent with them being true
steady states rather than transients; and the attractive-phoresis phases
(micellar, liposomal, active glass) show progressively slower structural
rearrangement from micellar to active glass, consistent with the
increasingly arrested dynamics described in Sec.~\ref{sec:attractive}. Videos of these six identified phases are available in \cite{SupplementalMaterial}.\\

\section{Analysis methods}\label{app:methods}

\subsection{Velocity field, energy spectrum, enstrophy}\label{app:spectra}

Head positions/orientations are deposited onto an $N_g\times N_g$ grid
via a Gaussian kernel ($\sigma=1.5b$):
\begin{equation}
  \bm v(\bm r) = \frac{\sum_i \hat{\bm n}_i\, e^{-|\bm r-\bm r_i|^2/2\sigma^2}}
                      {\sum_i e^{-|\bm r-\bm r_i|^2/2\sigma^2}}.
\end{equation}
The energy and enstrophy spectra follow from the 2D FFT $\hat{\bm
v}(\bm k)$ of the Gaussian-deposited velocity field ($\sigma = 1.5b$),
radially averaged over all wavevectors of the same magnitude
$|\bm k|=k$:
\begin{equation}
  \Ek = \tfrac{1}{2}\avg{|\hat{\bm v}(\bm k)|^2}_{|\bm k|=k},
  \qquad
  \Ok = \tfrac{1}{2}\avg{|i\bm k\times\hat{\bm v}(\bm k)|^2}_{|\bm k|=k}.
\end{equation}
$\avg{\cdot}_{|\bm k|=k}$ in addition denotes averaging over
$10$ snapshots evenly spaced across the final $30\%$ of each
trajectory once the system has reached steady state. $\omega = \partial_x v_y - \partial_y v_x$ on the same grid.
Turbulence identification relies on $k^*$ (the $\Ok$ peak), the
vorticity field, and $\Cv$ below.

\subsection{Velocity correlation \texorpdfstring{$\Cv(r)$}{Cv(r)}}

\begin{equation}
  \Cv(r) = \frac{\avg{\bm v(\bm r_0)\cdot\bm v(\bm r_0+\bm r)}_{|\bm r|=r}}
                {\avg{\bm v(\bm r_0)\cdot\bm v(\bm r_0)}},
\end{equation}
steady-state and radially averaged. A negative minimum signals
anti-correlated flow on either side of a vortex core (Sec.~\ref{sec:repulsive}).
\subsection{Clustering and percolation}\label{app:methods:clustering}

The adjacency graph is built from head positions only (tail monomers
are excluded from this diagnostic entirely): two heads are connected
by an edge if their separation is within $r_{\rm cut}=3b$. Clustering
coefficient \cite{Newman2003}:
\begin{equation}
  C = \frac{1}{N}\sum_i \frac{2 e_i}{k_i(k_i-1)},
\end{equation}
where $N$ is the number of heads, $k_i$ is the degree of head $i$ in
this head-only graph, and $e_i$ is the number of edges among head
$i$'s neighbours.
Percolation diagnostics: $f_{\max} = \max_c |c| / N$ (largest connected
component fraction) and $n_{\rm clusters}$ = number of components.
$C$ cannot separate many small dense aggregates from one large one;
$f_{\max}, n_{\rm clusters}$ can (Fig.~\ref{fig:attractive}d).

\subsection{Pair correlation and structure factor}
We study the 
spatial
pair
correlation
function, which is
defined
as:
\begin{equation}
  g(r) = \frac{1}{\rho N}\Big\langle\sum_{i\neq j}\delta(r - |\bm r_i-\bm r_j|)\Big\rangle
  \Big/ 2\pi r,
\end{equation}
periodic minimum-image, $\rho$ = head number density. $g(\theta)$: histogram
of bond angles $\arg(\bm r_i - \bm r_j)$ for pairs within $r_{\rm cut}=3b$,
folded to $[0^\circ,60^\circ)$ (hexatic test, Fig.~\ref{fig:attractive}e).\\

The structure factor is defined as the Fourier transform of the pair
correlation:
\begin{equation}
  S(\bm q) = 1 + \rho\int d^2r\;[g(r)-1]\,e^{-i\bm q\cdot\bm r}.
\end{equation}
Since $g(r)$ depends only on separation, not direction, the angular
part of this integral can be performed exactly, reducing $S(\bm q)$ to
a function of $q=|\bm q|$ alone via the 2D isotropic (Hankel) form:

\begin{equation}
  S(q) = 1 + 2\pi\rho\int_0^\infty r\,[g(r)-1]\,J_0(qr)\,dr,
  \qquad
  S_\theta(k) = \Big|\sum_\theta g(\theta)\,e^{-2\pi i k\theta/60^\circ}\Big|,
\end{equation}
where $J_0$ is the zeroth-order Bessel function of the first kind (not
to be confused with the phoretic source strength $\Jzero$ used
elsewhere in this paper); $k\geq1$ (Fig.~\ref{fig:attractive}f). $S(q)$
is a transform of the measured $g(r)$, not an independent measurement.

\subsection{Mean-squared displacements}
The mean squared displacement (e.g in Fig. \ref{fig:flock}) is defined as:
\begin{equation}
  \avg{(\Delta r)^2}(\Delta t) = \avg{|\bm r_i(t_0+\Delta t) - \bm r_i(t_0)|^2}_{i,t_0},
\end{equation}
where the average is taken over particles in the system, and positions are unwrapped across periodic boundaries before differencing.

\subsection{Number fluctuations}
The number fluctuations $\Delta N$ (c.f. Fig. \ref{fig:flock}) is given by
\begin{equation}
  \Delta N(\avg{N}) \sim \avg{N}^{\alpha},
\end{equation}
fit by linear regression in $\log\Delta N$ vs. $\log\avg{N}$, pooled
across densities, over a sequence of coarsening grids. $\alpha=1/2$
is the Poissonian reference.

\section{Simulation details}
Particle positions and orientations evolve according to the dynamical equations in Eq. (1) using a forward Euler–Maruyama integration scheme with a 
time-step that
 satisfies a stability criterion $\Delta t \leq 0.02\bead/v_{\mathrm max}$
The system employs periodic boundary conditions along both the $x$- and $y$-axes. Chemical interactions are evaluated using the minimum image convention, with selected parameter sets cross-validated against full Ewald summation.
Unless stated otherwise: $\ksp = \kbend = 175\bead$,
self-propulsion speed $\vs = 2.5\bead/\tau$,
$|\Jzero| = 1$, $D_t = 0$, $D_r = 0$, $\Lx = 100\bead$.
Area fraction $\phiA$ ranges from $0.05$ to $0.65$ and
$\Npc \in \{2, 4, 6, 8, 10, 12\}$.
The phoretic couplings are set to $\chiT = \chiR = 1$ throughout, except for Fig, \ref{fig:turbulent} where $\chiT = 8$, $\chiR = 4$ is chosen. Varying their ratio would interpolate between purely translational
and purely rotational phoretic response, which we leave for future work.
Positional and rotational fluctuations (noise terms) are intentionally kept sub-dominant relative to deterministic chemical interactions and self-propulsion.  Thus, we seek to uncover emergent dynamics 
entirely through deterministic mechanisms.

\section{Coarse-grained hydrodynamic theory: derivation details}\label{app:theory}

This section gives the full derivation underlying the kinetic theory
summarized in Sec.~\ref{sec:theory}, extending it from the one-body
conservation law through to the linear stability analysis.

\subsection{Microscopic equations of motion and reduction to the one-body level}

We model each chain of $\Npc$ monomers, in the rigid-rod limit, as a
single effective degree of freedom: a center-of-mass position $\bm R$
and orientation $\theta$, with $\bm R$ standing in for the collective
translational coordinate of the whole rod rather than any single
monomer. Physically, this assumes the chain's internal (spring) forces
equilibrate fast enough that the rod moves as a rigid unit driven by
the net force and torque acting on it, so a single self-propulsion
speed $\vs$ and a single translational phoretic coupling $\chiT$ can be
assigned to the rod as a whole, even though at the microscopic
(bead-spring) level only the head monomer self-propels and every
monomer independently senses the phoretic field. The rotational
coupling $\chiR$, which enters through the tip-localised torque
$\Omega$ and sets the polar instability threshold derived below,
depends only on the geometry of the tip offset and is unaffected by
this coarse-graining of the translational degrees of freedom.

Each rigid rod $i$ obeys overdamped Langevin equations for its center
of mass $\bm{R}_i$ and orientation $\theta_i$:
\begin{align}
  \dot{\bm{R}}_i &= \vs\,\nhat_i
    + \mu_T\,\bm{F}_i^{\rm ev}
    + \chiT\,\bm{J}(\bm{r}_{\rm tip}^i)
    + \sqrt{2D_t}\,\xi_i(t), \label{eq:rdot_app} \\
  \dot{\theta}_i  &= \mu_R\,\tau_i^{\rm ev}
    + \chiR\,\bigl(\nhat_i \times \bm{J}(\bm{r}_{\rm tip}^i)\bigr)_z
    + \sqrt{2\Dr}\,\eta_i(t), \label{eq:thdot_app}
\end{align}
where $\bm{r}_{\rm tip}^i = \bm{R}_i + \tfrac{\ell}{2}\nhat_i$ is the
head-monomer position (Sec.~\ref{sec:theory}), $\bm{F}_i^{\rm ev}$ and
$\tau_i^{\rm ev}$ are excluded-volume forces/torques, and $\xi_i(t)$
and $\eta_i(t)$ are unit-variance white noise.
Though the simulations reported in this paper are noiseless, we retain
$\Dr$ and $D_t$ here as a formal coarse-graining device, with all
quoted results understood in the limit $\Dr,D_t \to 0$. Crucially, the phoretic flux is
evaluated at the \emph{tip}, not at the center of mass:
\begin{equation}\label{eq:Jflux_app}
  \bm{J}(\bm{r}_{\rm tip}^i)
  = \Jzero \sum_{j \neq i}
    \frac{\bm{r}_{\rm tip}^i - \bm{r}_{\rm tip}^j}
         {|\bm{r}_{\rm tip}^i - \bm{r}_{\rm tip}^j|^3},
\end{equation}
the gradient of the monopolar concentration field
$c_j(\bm{r}) = \Jzero/|\bm{r}-\bm{r}_{\rm tip}^j|$ sourced at every
other rod's tip. This tip-dependence, rather than any dependence on
$\bm{R}_i$ directly, is what ultimately makes the pair torque
$\Omega$ (below) sensitive to chain length: a point-particle
($\ell=0$) rod feels a phoretic flux identical in form but centerd on
its own center of mass, which we show below produces no net polar
torque by symmetry.

The joint $N$-body probability density
$\Psi(\bm{R}_1,\theta_1,\ldots,\bm{R}_N,\theta_N,t)$ evolves under
Eqs.~\eqref{eq:rdot_app}--\eqref{eq:thdot_app} by the corresponding
many-body Fokker--Planck (Smoluchowski) equation,
\begin{equation}\label{eq:FP_Nbody}
  \partial_t \Psi
  = -\sum_i \Bigl[
      \nabla_{\bm{R}_i}\cdot\bigl(\dot{\bm{R}}_i^{\rm det}\Psi\bigr)
      + \partial_{\theta_i}\bigl(\dot{\theta}_i^{\rm det}\Psi\bigr)
    \Bigr]
    + \sum_i \Bigl[
      D_t \nabla_{\bm{R}_i}^2 \Psi + D_r \partial_{\theta_i}^2\Psi
    \Bigr],
\end{equation}
where $\dot{\bm{R}}_i^{\rm det}$, $\dot{\theta}_i^{\rm det}$ denote the
deterministic (non-noise) parts of Eqs.~\eqref{eq:rdot_app}--\eqref{eq:thdot_app}.
Integrating Eq.~\eqref{eq:FP_Nbody} over all but one rod's coordinates
defines the one-body marginal
$\psi(\bm{R},\theta,t) = N\int \Psi\,d\bm{R}_2 d\theta_2 \cdots d\bm{R}_N d\theta_N$,
which obeys an \emph{unclosed} equation involving the two-body
marginal $\psi_2(\bm{R},\theta,\bm{R}',\theta',t)$ through the
pairwise excluded-volume and phoretic terms. Closing this hierarchy at
the pair level via molecular chaos,
\begin{equation}\label{eq:molchaos}
  \psi_2(\bm{R},\theta,\bm{R}',\theta',t)
  \approx \psi(\bm{R},\theta,t)\,\psi(\bm{R}',\theta',t),
\end{equation}
yields the closed one-body conservation law used throughout the rest
of this appendix,
\begin{equation}\label{eq:cons_app}
  \partial_t \psi + \nabla_{\bm{R}} \cdot \bm{J} + \partial_\theta J^\theta = 0,
\end{equation}
with currents
\begin{align}
  \bm{J}(\bm{R},\theta,t)
  &= \vs\,\nhat\,\psi
    - D_t \nabla_{\bm{R}}\psi
    + \bm{J}^{\rm ev}[\psi]
    + \bm{J}^{\rm ph}[\psi], \label{eq:Jtr_app}\\
  J^\theta(\bm{R},\theta,t)
  &= - D_r \,\partial_\theta\psi
    + J^{\theta,\rm ev}[\psi]
    + J^{\theta,\rm ph}[\psi], \label{eq:Jrot_app}
\end{align}
whose explicit mean-field forms (obtained by substituting
Eq.~\eqref{eq:molchaos} into the pairwise terms inherited from
Eqs.~\eqref{eq:rdot_app}--\eqref{eq:Jflux_app}) are given in the next
two subsections. The translational diffusivities along and
perpendicular to the rod axis are $D_\parallel=2D_\perp\equiv D_t$ in
the dilute limit \cite{DoiEdwards}.



\subsection{Excluded-volume currents}

Following \cite{BM2008}, the excluded-volume currents are written in
the mean-field (molecular chaos) approximation as
\begin{align}
  \bm{J}^{\rm ev}[\psi](\bm{R},\theta)
  &= \mu_T \int d\bm{R}'\int d\theta'\;
    \bm{F}^{\rm ev}(\bm{R}-\bm{R}',\theta,\theta')\;
    \psi(\bm{R},\theta)\,\psi(\bm{R}',\theta'), \label{eq:Jev_ev_app}\\
  J^{\theta,\rm ev}[\psi](\bm{R},\theta)
  &= \mu_R \int d\bm{R}'\int d\theta'\;
    \tau^{\rm ev}(\bm{R}-\bm{R}',\theta,\theta')\;
    \psi(\bm{R},\theta)\,\psi(\bm{R}',\theta'). \label{eq:Jrot_ev_app}
\end{align}
For hard rods, $\bm{F}^{\rm ev}$ is a contact force active only when
two rods overlap, with effective cross-section
$\sigma(\theta-\theta') = \ell b |\sin(\theta-\theta')|$ \cite{Onsager}.
The detailed form of the collision integral is given in \cite{BM2008};
we do not reproduce it here since the excluded-volume interaction does
not by itself produce polar order for rigid rods \cite{BM2008} and its
main effect is to renormalise the translational diffusivity and to set
the hard-core domain entering $K_1(\ell)$ below.

\subsection{Phoretic currents}

The phoretic active currents are the key new ingredient. In the
mean-field approximation, the phoretic flux at the tip of rod 1 due to
rod 2 is
\begin{equation}\label{eq:Jpair_app}
  \bm{v}^{\rm ph}(\bm{R}_1,\theta_1;\bm{R}_2,\theta_2)
  = \chiT \Jzero\,
    \frac{\bm{r}_{\rm tip}^1 - \bm{r}_{\rm tip}^2}
         {|\bm{r}_{\rm tip}^1 - \bm{r}_{\rm tip}^2|^3},
\end{equation}
where $\bm{r}_{\rm tip}^k = \bm{R}_k + \tfrac{\ell}{2}\nhat_k$
(Sec.~\ref{sec:theory}). Introducing the tip-to-tip separation
$\bm{\Delta}_{12} = \xivec + \tfrac{\ell}{2}(\nhat_1-\nhat_2)$,
$\xivec = \bm{R}_1-\bm{R}_2$, the phoretic translational and rotational
currents are
\begin{align}
  \bm{J}^{\rm ph}[\psi](\bm{R},\theta)
  &= \chiT\Jzero \int d\bm{R}'\int d\theta'\;
    \frac{\bm{\Delta}_{12}}{|\bm{\Delta}_{12}|^3}\;
    \psi(\bm{R},\theta)\,\psi(\bm{R}',\theta'), \label{eq:Jph_app}\\
  J^{\theta,\rm ph}[\psi](\bm{R},\theta)
  &= \int d\bm{R}'\int d\theta'\;
    \Omega(\xivec,\theta,\theta';\ell)\;
    \psi(\bm{R},\theta)\,\psi(\bm{R}',\theta'), \label{eq:Jrph_app}
\end{align}
with pair torque
\begin{equation}\label{eq:Omega_app}
  \Omega(\xivec,\theta_1,\theta_2;\ell)
  = \chiR\Jzero\,
    \frac{(\nhat_1 \times \xivec)_z
         + \tfrac{\ell}{2}\sin(\theta_1-\theta_2)}
         {|\xivec + \tfrac{\ell}{2}(\nhat_1-\nhat_2)|^3}.
\end{equation}
The numerator has two contributions: (i) $(\nhat_1\times\xivec)_z$, a
torque due to the center-of-mass separation, identical to the
point-particle interaction; and (ii) $\tfrac{\ell}{2}\sin(\theta_1-\theta_2)$,
a purely geometric term arising from the tip offset, which is new and
carries the rod-length dependence.

\subsection{Moment equations and the polar instability threshold}

Expanding $\psi$ in angular Fourier modes and taking the zeroth and
first moments of Eq.~\eqref{eq:cons_app} under molecular chaos gives
coupled equations for the density $\rho$ and polarisation $\bm{P}$; the
polarisation equation takes the form
\begin{equation}\label{eq:P_app}
 \partial_t\bm{P} = -a_1(\rho_0)\,\bm{P} + \ldots,
  \qquad
  a_1(\rho_0) = \frac{K_1(\ell)\,\rho_0}{2} + \Dr,
\end{equation}
where
\begin{equation}\label{eq:K1int_app}
  K_1(\ell)
  = \frac{1}{\pi}\int_0^{2\pi}d(\Delta\theta)\sin(\Delta\theta)
    \int_{\mathcal D(\Delta\theta)}
    d^2\xivec\;
    \Omega(\xivec,\Delta\theta;\ell),
\end{equation}
$\mathcal D(\Delta\theta) = \{\xivec: d_{\min}(\xivec,\Delta\theta;\ell)
\geq 2b\}$ is the hard-rod excluded-volume domain (Sec.~\ref{sec:theory},
Eq.~\eqref{eq:dmin_main}), and $K_1(\ell)$ is normalised consistently with
the main text (this matches Eq.~\eqref{eq:Pgrowth_main} exactly; an
earlier working version of this note carried an additional spurious
factor of $2\pi$ in $a_1$, now removed). The instability condition is
$a_1 < 0$, i.e.
\begin{equation}\label{eq:instab_app}
  \rho_0 > \rho_c(\ell) = \frac{2\Dr}{|K_1(\ell)|}, \qquad \text{provided } K_1(\ell) < 0,
\end{equation}
consistent with Sec.~\ref{sec:theory}: $K_1<0$ is polar-aligning,
$K_1>0$ is anti-aligning and no homogeneous polar instability exists at
any density.

\subsection{Evaluation of \texorpdfstring{$K_1(\ell)$}{K1(l)}}

$K_1(\ell)$ is evaluated by direct numerical quadrature of
Eq.~\eqref{eq:K1int_app}, using an exact segment-segment distance
routine for the hard-rod domain $\mathcal D(\Delta\theta)$ (a naive
clamped-coordinate approximation to this distance was found to
substantially misestimate $d_{\min}$ in some configurations, though its
effect on the converged $K_1(\ell)$ values proved small, $\lesssim 3\%$).
Since $\Omega \sim 1/\xi^2$ at large $\xi$, the radial integral is only
conditionally convergent and relies on angular cancellation, so we
evaluate it over a symmetric $\xi$-domain rather than an arbitrary
cutoff.

\subsection{Linear stability at finite \texorpdfstring{$k$}{k} (heuristic)}

Perturbing the homogeneous state and Fourier transforming in space, the
transverse (purely orientational) mode disperses as
\begin{equation}\label{eq:sigma_T_app}
  \sigma_T(k) = -\Dr - D_t k^2 - \frac{\rho_0 K_1(\ell)}{2} + \ldots,
\end{equation}
recovering the $k=0$ instability threshold above. A finite-$k$
extension of this dispersion relation, and the resulting vortex-scale
estimate $k^* \sim (\rho_0|K_1(\ell)|/D_t)^{1/2}$ quoted in the main
text, would require a proper gradient expansion of the pair-torque
kernel $\Omega(\xivec,\Delta\theta;\ell)$ beyond the $k=0$ mode
retained here; this has not been carried out, and $k^*$ should be
understood as a heuristic estimate of the instability scale rather
than a derived fastest-growing-mode wavenumber (see Sec.~\ref{sec:theory}
for the corresponding caveat in the main text).

\subsection{Summary}

The framework above --- one-body kinetic equation, phoretic pair
torque, polar alignment coefficient $K_1(\ell)$, and the $k=0$
instability threshold --- is on solid footing and reproduces the
qualitative sign structure reported in Sec.~\ref{sec:theory}
($K_1<0$ for repulsive phoresis, $K_1>0$ for attractive). The
quantitative $\Npc$-dependence of the instability threshold, however,
is not currently explained by this theory at the rigid-rod, dilute,
mean-field level: $|K_1(\ell)|$ increases rather than decreases with
$\Npc$, so the theory alone does not account for the observed
suppression of flocking at $\Npc>2$. As discussed in the main text,
we attribute this to physics not included here --- most plausibly the
excluded-volume interaction of a genuinely semi-flexible chain, rather
than an idealised rigid rod, and finite-density pair correlations
beyond the molecular chaos approximation.


\bibliography{refs_chains}

\end{document}